
\tolerance=1200
\magnification 1200
\baselineskip=12pt plus 1pt
\parindent=25pt
\font\rmb=cmr10 scaled 1600

\font\small=cmr10 at 10truept
\baselineskip=20pt plus 1pt


{\nopagenumbers
{
\small
\baselineskip=12pt plus 1pt
\hfill Alberta-THY-14-1993

\hfill MARCH 1993
}

\vskip1cm
{\bf
\centerline{\rmb Exact Primordial Black Strings In Four Dimensions}
}

\vskip1.5cm

\centerline{\bf Nemanja Kaloper}
\centerline{Theoretical Physics Institute}
\centerline{Department of Physics, University of Alberta}
\centerline{Edmonton, Alberta T6G 2J1, Canada }
\centerline{email: kaloper@fermi.phys.ualberta.ca}
\vskip1.5cm
\centerline{\bf Abstract}

{
A solution of effective string theory in four dimensions is presented
which admits interpretation of a rotating black cosmic string.
It is constructed by tensoring the three
dimensional black hole, extended with the
Kalb-Ramond axion, with a flat direction. The physical interpretation
of the solution is discussed, with special attention on the axion, which
is found to play a role very similar to a Higgs field.
Finally, it is pointed out that the solution
represents an exact WZWN $\sigma$ model on the string world sheet,
to all orders in the inverse string tension $\alpha'$.
}
\vskip2cm
\centerline{\it Submitted to Phys. Rev. Lett.}
\vfil
\eject}

In recent years we have witnessed a very rapid growth of
the family of ``black'' configurations, representing
gravitational fields with event horizons of various
topology. They have ranged from various black holes [1-9],
over string-like configurations [1,10-13], to p-branes [7,13-14].

In this  letter, I will attempt to expand this family
by constructing a solution of the four dimensional
effective action of string theory
which admits the interpretation as a black cosmic
string inside a domain of axion field gradient.
The configuration is primordial, in the sense that the
domain is essential for its existence, because the ``axion charge''
(i.e., the gradient of the pseudoscalar axion) is what
stabilizes the string. Its
geometric structure is that of the three dimensional black hole recently
found by Banados, Teitelboim and Zanelli (BTZ)[3], tensored with a flat line
which is interpreted as the string axis. The solution could also be
viewed in light of toroidal black holes investigated by Geroch
and Hartle [15], and could be understood as such a black hole, which is bigger
than the cosmological horizon of the Universe in which it is imbedded.

The dynamics of the background field
formulation of string theory is defined with the
effective action which,
in the Einstein frame and to order $O(\alpha'^{0})$, is
$$
S~=~\int d^{4}x\sqrt{g}
\big({1 \over 2\kappa^{2}}R -
{1 \over 6} e^{-2\sqrt{2}\kappa \Phi}H_{\mu\nu\lambda}H^{\mu\nu\lambda}-
{1\over 2} \partial_{\mu}\Phi \partial^{\mu} \Phi +
\Lambda e^{\sqrt{2}\kappa \Phi} \big) \eqno(1)
$$

\noindent where $R$ is the Ricci scalar,
$~H_{\mu\nu\lambda}=\partial_{[\lambda}B_{\mu\nu]}~$
is the antisymmetric field strength associated with the Kalb-Ramond
field $~B_{\mu\nu}~$, $~\Phi~$
is the dilaton field
and $\Lambda$ the cosmological
constant.
The metric is of signature
$+2$, the Riemann tensor is defined according to
$R^{\mu}{}_{\nu\lambda\sigma} = \partial_{\lambda}\Gamma^{\mu}{}_{\nu\sigma} -
\ldots $, and the cosmological constant is defined with the opposite sign
from the more usual GR conventions:
$\Lambda > 0$ denotes a negative cosmological constant.
It has been included to represent the
central charge deficit. In the remainder of this paper, I will
work in the Planck mass units: $\kappa^2 = 1$.

Instead of writing out explicitly Einstein's equations, I will
work in the action, as this approach simplifies finding
the solutions. The background ans\" atz is that of a stationary
axially symmetric metric:
$$
ds^{2}= \mu^2(r) ~dr^{2}+G_{jk}(r)~dx^j dx^k + \eta^2(r) dz^2
\eqno(2)
$$

\noindent where the $2 \times 2$ matrix $G_{jk}(r)$ is of signature $0$ as
the metric (2) is Lorentzian and one of the coordinates $\{x^k\}$ is timelike.
The $z$ coordinate in (2) is noncompact,
whereas the spacelike $x^k$ is compact.
The ``lapse'' function $\mu^2$ is kept arbitrary as its variation in (1) yields
the constraint equation. The dilaton $\Phi$ is a function of $r$
only, and the axion equations
$\partial_{[\mu}H_{\nu\lambda\sigma]}=0$ and
$\partial_{\mu}\Bigl(\sqrt{g}\exp(-2\sqrt{2}\Phi)H^{\mu\lambda\sigma}\Big)=0$
are solved in terms of the dual vector field $V^{\mu}$
by
$$H_{\nu\lambda\sigma}=\exp{(2\sqrt{2}\Phi)}\sqrt{g}
\epsilon_{\mu\nu\lambda\sigma}V^{\mu}
\eqno(3)
$$

\noindent and $V=Q~dz$ (the topological charge term).
This solution has been discussed at more length
in [16] (see also [11]).
There, it has been argued that the axion equations of motion can be
solved by topological charge terms in cylindrical backgrounds, since
such topologies include a nontrivial first cohomology
from a non-contractible loop $S^1$ in the manifold. The
charge $Q$ above can therefore be thought of as associated with such a loop
of string, except that the string size is bigger than the cosmological horizon.

The simple form of the axion allows that it be integrated out from the
action, with the help of a Lagrange multiplier, and treating the solution (3)
as a constraint. The resulting effective action for the
dilaton-gravity system is
$$
S~=~\int d^{4}x\sqrt{g}
\big({1 \over 2}R -{1 \over2}\partial_{\mu}\Phi \partial^{\mu} \Phi
- {Q^2 \over \eta^2}~e^{2\sqrt{2}\Phi} +
\Lambda e^{\sqrt{2} \Phi} \big) \eqno(4)
$$

\noindent Note that the action (1) has been rewritten in the
form similar to Einstein gravity with a minimally coupled self-interacting
scalar field.
Also note that the sign of $Q^2$ is negative. This is
a consequence of the proper replacement of the axion with
its charge using the Lagrange multiplier method, and can be
verified from the inspection of the equations of motion derived from (1).
The action (4) will be discussed at more length later, with special
emphasis on the role of the axion in it.
The dilaton equation of motion is just
$$
\nabla^2 \Phi = {\partial V(\Phi) \over \partial \Phi}
\eqno(5)
$$

\noindent with the potential $V(\Phi)= (Q^2/\eta^2)~e^{2\sqrt{2}\Phi} -
\Lambda e^{\sqrt{2} \Phi}$. It is obviously desirable to look
for those solutions with the dilaton minimizing the potential $V(\Phi)$,
since if they exist, being the dilaton ``vacua'' they are the minimum
energy solutions of (1).
A slight complication is the presence
of $\eta$ in $V(\Phi)$. However, a solution can be sought for which
$\eta=1$. This should be done carefully because it must be
verified that $\eta=1$ represents a solution of the equations of motion.
An easy way to check it is as follows.

The background has three toroidal coordinates $\{x^k\}$
and $z$ which are
dynamically irrelevant. Hence the problem is
effectively one-dimensional and the
Kaluza-Klein reduction [17] can be employed to simplify  the action (1).
It is instructive here to perform the Kaluza-Klein reduction in two
steps, in order to isolate the dynamics of the mode $\eta$. The first
step is to integrate out the coordinate $z$. The resulting effective action
in three dimensions is, after the rescaling
$S^{3D}_{\rm eff} = 2 S/\int dz$,
$$
S^{3D}_{\rm eff}~=~\int d^3x\sqrt{G}
\big({1 \over 2} \eta \bar R -{1 \over2}
\eta \partial_{\mu}\Phi \partial^{\mu} \Phi
- {Q^2 \over \eta}~e^{2\sqrt{2}\Phi} +
\eta \Lambda e^{\sqrt{2} \Phi} \big) \eqno(6)
$$

\noindent The mode $\eta$ (the ``compacton'') in this form of the action is
obviously just a Lagrange multiplier. Its Euler-Lagrange equation however
involves the $3D$ part of the metric. Thus, to investigate it it is neccessary
to write out the complete set of Einstein's equations in addition to it. This
can be avoided with a conformal redefinition of the $3D$ metric such that
the $3D$ Ricci curvature disappears from the $\eta$ equation. The conformal
rescaling which ensures this is $G_{\mu\nu} = (1/\eta) \tilde G_{\mu\nu}$. The
resulting action is just
$$
S^{3D}_{\rm eff}~=~\int d^3x \sqrt{\tilde G} \big({1 \over 2} \tilde R
- {1 \over \eta^2} \partial_{\mu}\eta \partial^{\mu} \eta
- {1 \over2} \partial_{\mu}\Phi \partial^{\mu} \Phi
- {Q^2 \over \eta^4}~e^{2\sqrt{2}\Phi} +
{\Lambda \over \eta^2}~e^{\sqrt{2} \Phi} \big) \eqno(7)
$$

\noindent This action represents ordinary $3D$ General Relativity
with two minimally coupled self-interacting fields $\Phi$ and $\ln \eta$. The
effective potential is now
$$
V_{\rm eff}(\Phi, \eta)= {Q^2\over \eta^4}~e^{2\sqrt{2}\Phi} -
{\Lambda \over \eta^2}~e^{\sqrt{2} \Phi}
\eqno(8)
$$

\noindent and the equations of motion for the scalars are (5) (where
$V(\Phi)$ is replaced by $V_{\rm eff}(\Phi, \eta)$) and
$$
2 \nabla^2 \ln \eta =
{\partial V_{\rm eff}(\Phi, \eta) \over \partial \ln \eta}
\eqno(9)
$$

Interestingly, the system of equations (5), (9) is simultaneously solved
by a ``vacuum'' $\Phi=\Phi_0$, $\eta=\eta_0$ provided that
$$
Q^2 = {\Lambda \over 2}\eta_0^2~e^{-\sqrt{2} \Phi_0}
\eqno(10)
$$

\noindent This equation can always be satisfied, and actually can
be viewed as the definition of the dilaton vacuum expectation value given
the other parameters. The resulting effective action for gravity in
three dimensions under the
assumption that the ``matter'' modes
are in ``vacuum'' can be obtained upon substitution of (10) in (7). It is
$$
S^{3D}_{\rm eff~vac}~=~\int d^3x \sqrt{\tilde G}
\big({1 \over 2} \tilde R
 + \lambda_{3D}  \big) \eqno(11)
$$

\noindent and represents just the normal $3D$ Einstein-Hilbert action with
an effective (negative) cosmological constant
$\lambda_{3D} = (\Lambda / 2\eta_0^2)~e^{\sqrt{2} \Phi_0}$. Its unique black
hole solution is the BTZ solution, as shown in [3,9]. Therefore, the next step
of dimensionally reducing (11) to a one-dimensional problem can  be skipped,
and as a consequence, the metric part of the  black string solution
can be written as $ds^2 = ds^2_{BTZ} + dz^2$ after setting $\eta_0 =1$.
The complete rotating black string solution of (1) is thus
$$\eqalign{
&ds^{2} =  { d\rho^{2} \over \lambda_{3D} (\rho^2 - \rho^2_{+})} +
{}~R^2 (d\theta + N^{\theta} dt)^2
- {\rho^2 \over R^2} {\rho^2 - \rho^2_{+} \over \lambda_{3D}} dt^2 + dz^2 \cr
&~~~~~~~~~~~~~~~~~~~~~~~~~~~~~~~~~~~~H_{t \rho \theta} = {\rho \over Q} \cr
&~~~~~~~~~~~~~~~~~~~~~~~~~~~~~~~~~~~~~~\Phi = \Phi_0
}
\eqno(12)
$$

\noindent with $\rho_+^2 = M(1 - (J/M)^2)^{1/2}$,
$R^2 = (\sqrt{\lambda_{3D}}/2)\bigl(\rho^2 + M - \rho_+^2 \bigr)$ and
$N^{\theta} = - J /2R^2$, and
where the identity (10) has been used.
The physical black strings should
also satisfy the constraint $\mid J \mid \le M$.
If this were not fulfilled, one would end
up with a singular structure, manifest by
the appearance of closed timelike curves in
the manifold accessible to an external observer, crossing the
point $R =0$. Such a voyage has been investigated
in [11] for the spinless
case, and also in [10] for the vacuum. Moreover,
it has been
argued that, although the solution (12) does not have
curvature singularities ($R_{\mu\nu\lambda 3}=0$,
$R_{\mu\nu\lambda\sigma}=-\lambda_{3D}\bigl(
g_{\mu\lambda}g_{\nu\sigma}-g_{\mu\sigma}g_{\nu\lambda}\bigr)$),
they can develop if the metric is slightly perturbed by
a matter distribution [3].
Thus, the singularities are hidden by a horizon
if the spin is bounded above
by the mass. By analogy
with the BTZ black hole,
the solution with $J=M$ is understood as
the extremal black string, and $J=M=0$
as the vacuum. There is
a local correspondence between these
two cases, as discussed in [9].

The solution (12) describes a rotating black string as is rather
obvious from the metric. However, it is the axion which gives further clues
regarding the nature of the string. As was mentioned above, one way to think
of the solution is to imagine it as a loop of string with its length
parametrized by $z$, which is bigger than the cosmological horizon of the
universe where it is imbedded. In this sense, the solution represents
an explicit example of a toroidal black hole [15].
Such an interpretation obviously
puts limits on the validity of the approximations underlying the assumption
that the string is straight. A more interesting picture is obtained if one
retains the image of the string as infinitely long and straight. The dual
axion field strength $V=Qdz=da(z)$ can be integrated between any two
space-like ($t={\rm const}$) hypersurfaces $z_{1,2}={\rm const}$ to give
$a(z_2) - a(z_1) = Q \Delta z$. Therefore, the axion solution can be
understood as a constant gradient of the pseudoscalar axion field. As
$z_{1,2} \rightarrow \infty$, the axion diverges. But this is easy to explain:
it is merely a consequence of the assumption that the string is infinitely
long. In reality, one should expect some cut-off sufficiently far away along
the string. The situation is precisely analogous to that of the
electrostatic potential between the plates of a parallel plate capacitor in
ordinary electromagnetism. There, the cut-off occurs on the plates of the
capacitor, where the potential assumes constant values. The gradient is just
$\vec \nabla V = (\Delta V/\Delta L) \vec z$. This analogy shows that the
black string solution (12) should be viewed as a gravitational configuration
which arose inside a transitory region
separating two domains within which
the axion is constant, $a_1$ and $a_2$ respectively. The axion gradient
inside this region corresponds to the adiabatic change in the axion vacuum,
where the adiabatic approximation is better if
the transitory region (and hence the string)
is bigger. The configuration (12) then evidently needs
the domain of axionic gradient for
its existence (because the axion gradient stops the dilaton from rolling), and
thence can justifiably be labelled primordial.

The discussion of the previous paragraph
illustrates only one aspect of the importance of the
axion in obtaining the solution (12). Besides providing the
extra contribution to the dilaton-compacton self-interactions, the axion
also plays role of a Higgs field, which is evident from the steps leading from
Eq. (1) to Eq. (4). The axion condensate $Q^2$ in (4) breaks the
normal general covariance group $GL(3,1)$ of (1) down to $GL(2,1)$ which
is the invariance group of (4). It should be noted, though, that
the Higgs-like behaviour of the axion is purely topological; indeed, in the
$O(\alpha'^0)$ approximation, the axion has no self-interactions, and hence
no potential to minimize. Again, this is not really a surprise. The behavior
of the Peccei-Quinn (PQ)
axion has been found very much the same, and at tree-level
the PQ axion condensate was also purely topological. It was only after the
radiative corrections were included, that its
self-interaction potential arose.
Hence, to investigate the Higgs aspect of the axion further it would
be necessary to inspect higher order corrections to (1).

This programme could be best conducted via the Wess-Zumino-Witten-Novikov
(WZWN) $\sigma$ model approach [2].
Namely, it was demonstrated recently that the
BTZ solution can be obtained as either a non-gauged WZWN model on the
group $SL(2,R)/P$ or an extremaly  gauged WZWN model on the coset
$(SL(2,R) \times R) /(R\times P)$, where $P$ is a discrete group which
represents compactification of one of the space-like coordinates
to a circle [8-9]. In this light, the solution (12) is obviously obtained by
taking either of these two $\sigma$ models and simply tensoring them with
an additional flat direction, which will be the coordinate along the string.
Thus, specifically, (12) is an extremaly  gauged WZWN model on the coset
$(SL(2,R) \times R^2) /(R\times P)$.
Higher order corrections could now be investigated following the resummation
procedure established by Tseytlin [18] and by Bars and Sfetsos [19].
It turns out, that the black string configuration actually survives the
corrections, and appears to be an exact solution of string theory to all
orders in $\alpha'$. The only effect of the higher order $\alpha'$
corrections is finite renormalization of the parameters in (12), and
in particular, renormalization of the semiclassical expression for the
cosmological constant . The details will be presented elsewhere [20].

There still remains the problem of
stability of solution (12) under small
perturbations. Some indications can be obtained
by looking at the ``matter'' sector of
the effective action $S^{3D}_{\rm eff}$ (7), after the conformal rescaling.
The dilaton and the
compacton in (7) can be viewed as an $O(2)$ doublet, with
the effective potential (8) manifestly breaking $O(2)$. As a consequence,
the linear combination  $\sqrt{2} \Phi - \ln \eta$
of the dilaton and compacton picks up a mass term of order $\Lambda$, whereas
its orthogonal complement remains massless. It would have been
preferable if both the dilaton and the compacton became massive, because
their big masses would de facto decouple them and
improve the stability of the
solution (12). As is, the solution (12) could actually
be spoiled by perturbations of the massless mode, which
can accumulate exterior to the black hole, much like the Goldstone modes
present in global cosmic string backgrounds [21]. This remains to be
investigated further in the future.

In closing, it has been shown that
the serendipitous BTZ $3D$ black hole has simple generalizations to four
dimensions, where it can be interpreted as a primordial spinning black string,
which is singularity-free.
The most attractive generalization is where
it represents a vacuum solution of tree level string theory, where the dilaton
and compacton have been decoupled due to the axion charge. In this respect, the
axion plays the role of a Higgs field,
since it breaks down the invariance group
$GL(3,1)$ of the underlying $4D$ theory down to the invariance group $GL(2,1)$
of the resulting three dimensional effective action, and modifies the
effective scalar potential of the model leading to the previously mentioned
decoupling of the scalar modes. The dilaton of the configuration is
constant and thus (12) also represents a solution
of four dimensional Einstein gravity with a minimally coupled
$3$-form field strength (see also [22]).
Moreover, the solution represents an
exact WZWN $\sigma$ model on the worldsheet, and thence can be easily extended
to include higher order $\alpha'$ corrections, as I will show elsewhere [20].
In the end, these do not
affect the nature of the solution, and it remains a well behaved  singularity
free string configuration with a horizon.

\vskip1cm
{\bf Acknowledgements}
\vskip0.5cm
I would like to thank B. Campbell and V. Husain for
comments on the manuscript and helpful discussions
and G. Hayward
for pointing out Ref. [14] to me.
This work has been supported in part by
the Natural Science and Engineering Research Council of Canada.

\vskip1cm
{\bf References}
\vskip0.5cm
\item{[1]} for a recent review of stringy aspects of black holes and strings
see J.A. Harvey and A. Strominger, Fermilab preprint
EFI-92-41/hep-th/9209055, Sep 1992; A. Sen, Tata Inst. preprint
TIFR-TH-92-57/hep-th/9210050, Oct 1992; G.T. Horowitz, UCSB preprint
UCSBTH-92-32/hep-th/9210119, Oct 1992; and references therein.
\item{[2]} E. Witten, Phys. Rev. {\bf D44} (1991) 314.
\item{[3]} M. Banados, C. Teitelboim, and J. Zanelli, Phys. Rev. Lett.
{\bf 69} (1992) 1849; M. Banados et. al., IAS preprint HEP-92/81
/gr-qc/9302012, Jan. 1993; D. Cangemi et. al.,
MIT preprint CTP-2162, 1992/gr-qc/9211013.
\item{[4]} G.W. Gibbons, Nucl. Phys. {\bf B207} (1982) 337;
M.J. Bowick et. al., Phys. Rev. Lett. {\bf 61} (1988) 2823;
D. Garfinkle et. al.,
Phys. Rev. {\bf D43} (1991) 3140; A. Shapere et. al.,
Mod. Phys. Lett. {\bf A6} (1991) 2677.
\item{[5]} A. Sen, Phys. Rev. Lett {\bf 69} (1992) 1006.
\item{[6]} R. Kallosh et. al.,
Phys. Rev. {\bf D46} (1992) 5278.
\item{[7]} D. Gershon, Tel Aviv University preprint TAUP-1937-91,
Dec 1991/hep-th/9202005;  TAUP-2005-92/hep-th/9210160, Oct 1992.
\item{[8]} G.T. Horowitz and D.L. Welch,
UCSB preprint NSF-ITP-93-21/hep-th/9302126, Feb 1993.
\item{[9]} N. Kaloper, Univ. of Alberta
preprint Alberta-Thy-8-93/hep-th/9303007, Feb. 1993.
\item{[10]} J.H. Horne and G.T. Horowitz, Nucl. Phys. {\bf B368} (1992) 444.
\item{[11]} N. Kaloper, Univ. of Minnesota preprint UMN-TH-1024/92,
May 1992, in press in Phys. Rev. {\bf D}.
\item{[12]} A. Sen, Nucl.Phys. {\bf B388} (1992) 457; E. Raiten, Fermilab
preprint FERMILAB-PUB-91-338-T, Dec. 1991; S. Mahapatra, Inst. of Math. Sci.
Madras preprint IMSC-93-06/hep-th/9301125, Jan 1993.
\item{[13]} I. Bars and K. Sfetsos, USC preprint USC-92/HEP-B1, May
1992/hep-th/9301047; Phys. Rev. {\bf D46} (1992) 4510; USC preprint
USC-92-HEP-B3, Aug. 1992/hep-th/9208001;
K. Sfetsos, USC preprint USC-92/HEP-S1, June 1992.
\item{[14]} S.B. Giddings and A. Strominger, Phys. Rev. Lett. {\bf 67}
(1991) 2930; G.T. Horowitz and A. Strominger, Nucl. Phys. {\bf B350}
(1991) 197; A. Sen, Phys. Lett. {\bf B274} (1992) 34;
S.K. Kar et. al.,
IP Bhubaneswar preprint IP/BBSR/92-35/hep-th/9205062, May 1992;
S. Mahapatra, Mod. Phys. Lett. {\bf A7} (1992) 2999.
\item{[15]} R. Geroch and J. Hartle, J. Math. Phys. {\bf 23}(4) (1981) 680.
\item{[16]} E. Witten, Phys. Lett. {\bf B153} (1985) 243.
\item{[17]} for a reviw see E. Cremmer, ``Dimensional Reduction
in Field Theory and Hidden Symmetries in Extended Supergravity'',
pp 313, in {\bf Supergravity 81}, ed. S. Ferrara and J.G. Taylor,
Cambridge Univ. Press, Cambridge 1982.
\item{[18]} A. Tseytlin, Imperial Coll. preprint
Imperial/TP/92-93/10/hep-th/9301015, Dec. 1992;
CERN preprint CERN-TH-6804/93/hep-th/9302083, Feb. 1993.
\item{[19]} I. Bars and K. Sfetsos,
USC preprint USC-93/HEP-B1/hep-th/9301047, Jan. 1993.
\item{[20]} N. Kaloper, ``More Twisted Cosmologies In String Theory'',
Univ. of Alberta preprint Alberta-Thy-7-93/hep-th/9303007, in preparation.
\item{[21]} see e.g. A. Vilenkin, Phys. Rep. {\bf 121} (1985) 263.
\item{[22]} G. Grignani and G. Nardelli, U. Perugia preprint
DFUPG-59-1992/gr-qc/9211001, Nov. 1992.

\bye